\documentclass[prc,twocolumn,showpacs]{revtex4}
\usepackage{graphicx}
\usepackage{dcolumn}
\usepackage{bm}

\begin{document}

\title{Mean field study of structural changes in Pt isotopes  
with the Gogny interaction.}

\author{R. Rodr\'{\i}guez-Guzm\'an$^{1,2}$, P. Sarriguren$^{2}$, 
L.M. Robledo$^{3}$ and  J.E. Garc\'{\i}a-Ramos$^{1}$.}

\affiliation{
$^{1}$ Departamento de F\'isica Aplicada, Universidad 
de Huelva, 21071 Huelva, Spain. 
\\
$^{2}$ Instituto de Estructura de la Materia, CSIC, Serrano
123, E-28006 Madrid, Spain.
\\
$^{3}$ Departamento  de F\'{\i}sica Te\'orica C-XI,
Universidad Aut\'onoma de Madrid, 28049-Madrid, Spain.
}

\date{\today}

\begin{abstract}
The evolution of the nuclear shapes along the triaxial
landscape is studied in the Pt isotopic chain using the
selfconsistent Hartree-Fock-Bogoliubov approximation
based on the Gogny interaction. In addition to the 
parametrization D1S, the new incarnations D1N and D1M
of this force are also included in our analysis 
to asses to which extent the predictions are independent 
of details of the effective interaction. The considered
range of neutron numbers $88 \le N \le 126$ 
includes prolate, triaxial, oblate and
spherical ground state shapes and serves for 
a detailed comparison of the predictions  
obtained with the new sets D1N and D1M  against the ones
provided by the standard parametrization Gogny-D1S
in a region of the nuclear landscape for which
experimental and theoretical fingerprints of 
shape transitions have been found. Structural 
evolution along the Pt chain is discussed in terms
of the deformation dependence of  single particle 
energies.

\end{abstract}

\pacs{21.60.Jz, 27.70.+q, 27.80.+w}

\maketitle

\section{Introduction.}
Quadrupole collectivity is one of the most relevant features of nuclear
structure \cite{Bo-M,rs}. In this context, the theoretical  understanding
of the evolution of the nuclear shapes, and the 
structural changes associated with it, represent 
an active research field
\cite{review,review-Bender,Heenen-nature,Werner,Robledo-1,Robledo-2,Rayner-1}. 
From the experimental side, low-lying 
 spectroscopy is one of the most
powerful sources of information about nuclear shapes and/or shape 
transitions since one can establish signatures correlating the 
excitation energies with the deformation properties 
\cite{Julin,draco,davidson,wu96,podolyak,pfun,caamano}.

Nowhere, however, the  evolution of nuclear shapes 
is more documented and challenging than 
around the proton shell closure $Z=82$. For example, the 
neutron deficient Lead isotopes with neutron number
$N \approx 104$ display three $0^{+}$ states within 
1 MeV excitation energy \cite{Andreyev-1}. The 
very rich and challenging variety of nuclear
shapes also extends to the neighboring Hg and Po isotopes
\cite{Julin}. In 
particular, it has been demonstrated 
\cite{delta2p-1} that in the case of the Lead isotopes
the 
decreasing 
trend observed in the 
binding energy difference 
${\delta}_{2p}(Z,N)=E(Z-2,N)-2E(Z,N)+E(Z+2,N)$
for decreasing mass number $A$, can already
 be described quantitatively  by mean field models
in terms of deformed ground states of Hg and Po nuclei
while the inclusion of  the quadrupole correlation energy
\cite{delta2p-2} brings the calculations even closer to 
experiment. From the experimental point of view 
\cite{Julin}, the neutron deficient Mercury isotopes
exhibit deformed  ground 
states while the situation is more involved in 
the case of Po nuclei (see, for 
example \cite{Grahn}, and references
therein).

A considerable effort, has also 
been devoted to 
characterize Pt nuclei \cite{davidson,wu96,King-exp,Cederwall-exp,
Seweryniak-exp,Kondev-exp,Blanc,Joss-exp,Soramel-exp,Popescu-exp,
Xu-exp}. In this case, several deformation regimes 
have been suggested. Previous theoretical
investigations \cite{Sauvage,Ansari,Krieger,bengtsson,Moller-1995,Moller-2008}
have found  triaxial and oblate 
ground state shapes
for the heaviest Pt isotopes 
while for the light ones a
prolate deformed regime 
is predicted. 
From the experimental point of view 
\cite{podolyak,caamano,CDROM}
the energy ratio $E_{4^{+}}/E_{2^{+}}$  
is almost  2.5 
for Pt nuclei with neutron number 
$110 \le N \le 118$ already pointing 
to  $\gamma$ soft shapes.
The role played by  the $\gamma$ degree of freedom 
in Pt isotopes has also been  stressed 
by the comparison
of experimental and theoretical results
performed in Ref. \cite{Hilberath}
which shows, that good agreement can 
be obtained if triaxiality is taken into account.
Further down, a transition
to a vibrational regime is suggested for $^{168-172}$Pt
by both the experimental data and their theoretical interpretation
\cite{Cederwall-exp}.

The shape evolution 
provided by the mean field 
framework \cite{rs}, based on the most recent 
incarnations of the Gogny interaction \cite{gogny}, in the 
isotopic chain
$^{166-204}$Pt is  considered in the present
study as a representative sample
of nuclei close to the $Z=82$ proton shell closure
for which prolate, triaxial, oblate and spherical
ground state shapes are found.

Nuclear shapes 
around the proton magic number  $Z=82$ have been studied using
a wide variety of theoretical
models. Low-lying minima 
in both Pb and Hg isotopes, have been predicted
within the framework of the Strutinsky method 
\cite{Pashkevich,Be-Na,Na-other}. From a mean field perspective, the 
coexistence between different nuclear shapes 
in Pt, Hg and Pb nuclei, has been considered
within the relativistic mean field approximation
\cite{Yoshida-1,Yoshida-2}. Deformed ground states
were predicted in the case of Pb isotopes as well as superdeformed
ground states in Hg isotopes at variance with
experiment. In order to cure this problem, a new
parametrization of the relativistic mean field Lagrangian, called 
NLSC, was introduced in Ref. \cite{NLSC}. More recently,  
a new set called NL3$^{*}$ has been proposed in Ref. \cite{NL3-star}
providing an improved description of the ground state properties of many nuclei 
like Pb isotopes. Studies based on Skyrme-like
models have been reported 
\cite{Valor,Meyer,Smirnova,Sarri-Moreno,Sarri-Moreno-other} while
 the shape coexistence in 
$^{182-192}$Pb was analyzed in Ref. \cite{rayner-PRL} using 
the Hartree-Fock-Bogoliubov (HFB) approach  based on the 
parametrization D1S \cite{gogny-d1s}
of the Gogny interaction \cite{gogny}. From a beyond mean 
field perspective, symmetry projected configuration mixing, based
on both Skyrme \cite{Duguet-1,Duguet-2} and Gogny 
\cite{Rayner-Pb}
energy density 
functionals, has been successfully employed in this region 
of the nuclear chart establishing a firm ground to support
the experimental evidences 
for rotational bands in the neutron deficient Pb isotopes
built, on coexisting low-lying $0^{+}$ states.  
On the other hand, evidences for $\gamma$ vibrations and shape
evolution have also been considered in the nuclei
$^{184-190}$Hg  \cite{delaroche94}, where a five-dimension collective 
Hamiltonian was built with the help of constrained Gogny-D1S
HFB calculations.
 
Just below the proton magic number $Z=82$, the nuclei with 
$A=170-200$ are particularly interesting because, small islands of 
oblate deformation might be favored energetically. Transitions from 
prolate to oblate shapes, as the number of neutrons 
increases, have been predicted 
in this mass region, using collective models 
\cite{wu96}, phenomenological Woods-Saxon or Nilsson potentials 
\cite{bengtsson,naza90,wheldon}, relativistic mean field 
\cite{naik,fossion,Sharma-Ring} as well as non-relativistic  
deformed Hartree-Fock (HF) \cite{stevenson,Robledo-2}
and HFB calculations \cite{Robledo-2,ours2}. Signatures
 for a transition from prolate to oblate
ground states, as the number of neutrons increases
from $N=110$ to $N=122$, have been found in Hf, W and Os isotopes. In 
particular, such a transition was found \cite{Robledo-2} to happen at 
$N=116-118$. Subsequently, the evolution of the ground state shapes
along the triaxial landscape of several isotopes of 
Yb, Hf, W, Os and Pt has been studied in Ref. \cite{ours2} within
the framework of the mean field approximation based on both 
Skyrme and Gogny interactions. This region is also an active research field  
within the Interacting Boson Model (IBM)
\cite{Barret-1,pairs-IBM-1,pairs-IBM-2,pairs-IBM-3,pairs-IBM-4,pairs-IBM-5,pairs-IBM-6,gramos,morales}.

Taking into account that
around the neutron mid-shell N=104  examples of
 coexisting configurations
have been found \cite{Julin}, it is very interesting to study 
the propagation of 
the nuclear shapes in the Pt isotopic chain
 and the way it can be correlated with 
the details in the underlying  single-particle levels as functions
of the deformation parameters. For such a task, the 
mean field approximation  appears as a first  tool 
incorporating, important correlations within the concept of spontaneous 
symmetry breaking and allowing  a  description of the evolution 
of shell structure and  deformation all over the 
nuclear chart 
(see, for example, \cite{review-Bender,Robledo-1,Robledo-2,Rayner-1,ours2,mf-1}
and references therein).

%%%%%%%%%%%%%%%%%%%%%%%%%%%%Fig1%%%%%%%%%%%%%%%%%%%%%%%%%%%%%%%%%%%%%%%%%%%%%%%%%%%%%%%%%
\begin{figure*}
\includegraphics[width=0.98\textwidth]{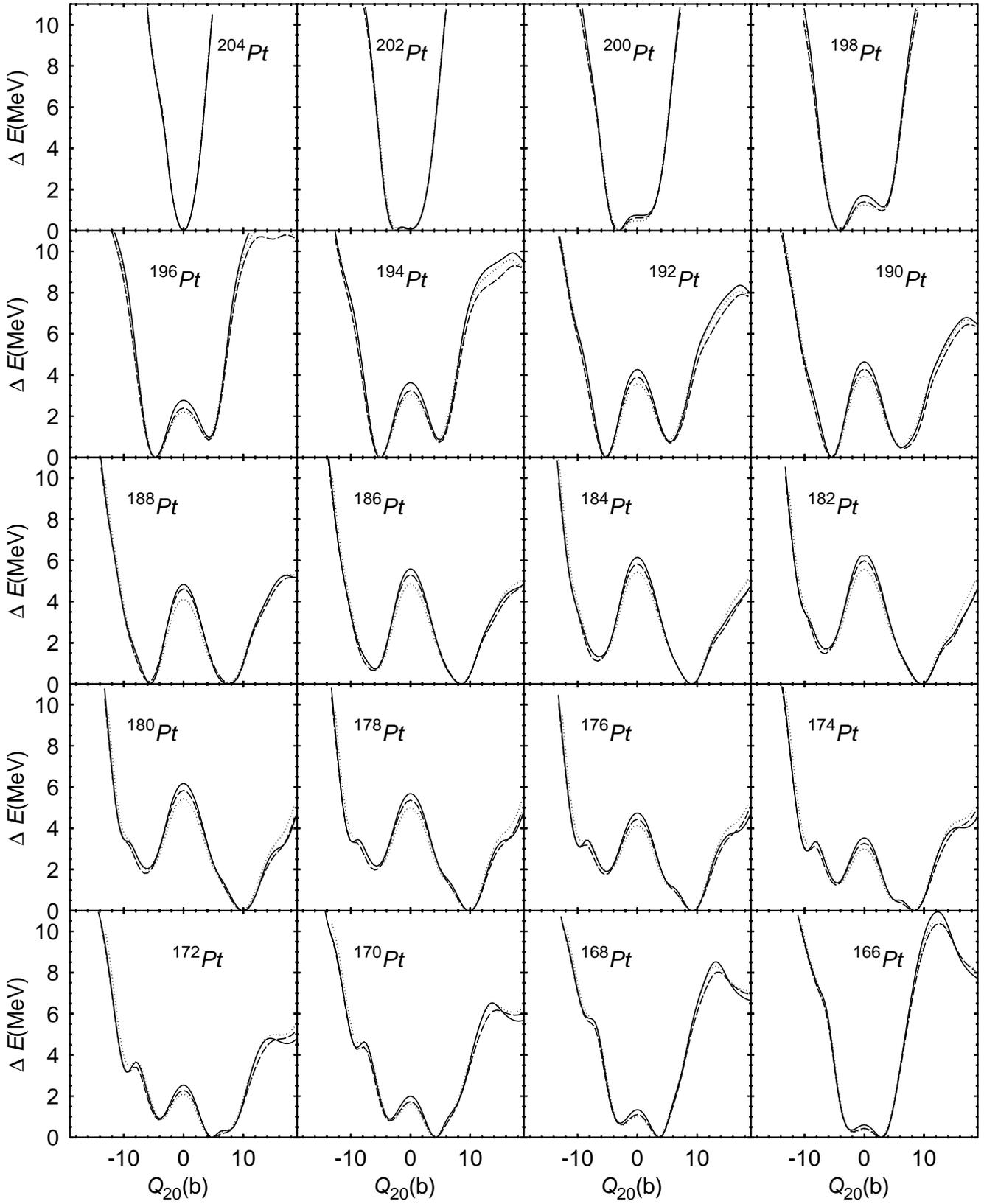}
\caption{Potential energy curves for the  isotopes
$^{166-204}$Pt as functions  of the axial 
quadrupole moment $Q_{20}$ calculated with the 
parametrizations D1S (continuous line), D1N (dashed line)
and D1M (dotted line)
of the Gogny interaction.}
\label{fig_axial}
\end{figure*}

In the present work, our 
study will be performed within the  
selfconsistent HFB
framework based on  the 
Gogny interaction \cite{gogny}. In addition to  D1S \cite{gogny-d1s}, which is 
still the most standard and thoroughly tested parametrization, we also
considered the two most recent parameter sets of the Gogny interaction, i.e.,
D1N \cite{gogny-d1n} and D1M \cite{gogny-d1m}. To the best of our 
knowledge, the results
to be discussed later on in this paper, are the first systematic 
mean field study reported in the 
literature, using
 both D1N and D1M parametrizations, to describe
 (mean field)
ground state properties of Pt nuclei. 
The selected
isotopes $^{166-204}$Pt cover almost the whole 
shell 
(i.e., $N=88-126$) and 
display a range of ground state
shapes  wide enough  so as to be considered a 
 very challenging testing ground 
 for a  comparison of the (mean field) ground state
 properties predicted with the new parametrizations 
 D1N   and D1M    against the 
 standard D1S    Gogny functional. Our calculations also
 included the isotopes $^{160-164}$Pt which  turn
 out to be spherical and  will
 not be further discussed in the present study. 
%The two
%older Gogny parametrizations D1 and D1$^{'}$ \cite{gogny} are  
%included in our analysis.

The paper is organized as follows. In Sec. \ref{formalism} we present a 
brief description of the theoretical framework
used. The 
results of our study are discussed 
in Sec. \ref{results}. There, we discuss, in Sec. \ref{results-PECPES}, our 
Gogny-HFB 
calculations, for which axial symmetry is preserved as selfconsistent 
symmetry, used to construct Potential Energy Curves (PECs). In a
 second step we discuss our study of the triaxial
 landscape, providing Potential Energy Surfaces (PESs), by 
 constraining on both $\beta$ and $\gamma$ 
quadrupole deformations. In our  axial and triaxial HFB 
calculations we considered at the same time, the three parameter sets 
D1S, D1N and D1M. The interaction Gogny-D1S, taken as a reference
in the present study, is already considered as a global force able to describe reasonably well 
low energy experimental data all over the nuclear chart 
(see, for example, \cite{rayner-PRL,gogny,gogny-d1s,
gogny-other-1,gogny-other-2,
gogny-other-3,gogny-other-4,gogny-other-5,
gogny-other-6,Robledo-1,Rayner-Pb,bertsch,
peru,gogny-other-7,gogny-other-8,hilaire,delaroche10} and references therein).
This is also likely to be the situation with the new Gogny interactions D1N 
\cite{gogny-d1n,ours2}  and D1M \cite{gogny-d1n} 
but still  further explorations are required. Therefore, we consider
all these interactions/functionals in the present study 
to asses to which extent the fine details of our mean field 
predictions for the nuclei $^{166-204}$Pt are independent 
of the particular version of the Gogny force employed.
After discussing the mean field
systematics of deformation for the considered nuclei, we
turn our attention, in Sec. \ref{results-SPE}, to the 
underlying single-particle properties as functions of both 
deformation (axial and triaxial) and mass number. This is 
relevant if one  keeps in mind that, from a mean field
perspective, shape changes arise when the deformed single-particle levels 
are energetically favored to a different degree in open shell nuclei
(Jahn-Teller effect \cite{JTE-1}). We also consider the 
behaviour with neutron number of the spherical shell occupancies corresponding
to the ground state of the different Pt isotopes  in order to
shed some light into the phenomena involved in the different deformation regimes.
Finally, Sec. \ref{conclusions} contains the concluding remarks and work perspectives.

%%%%%%%%%%%%%%%%%%%%%%%%%%%%Fig2%%%%%%%%%%%%%%%%%%%%%%%%%%%%%%%%%%%%%%%%%%%%%%%%%%%%%%%%%
\begin{figure*}
\includegraphics[width=0.98\textwidth]{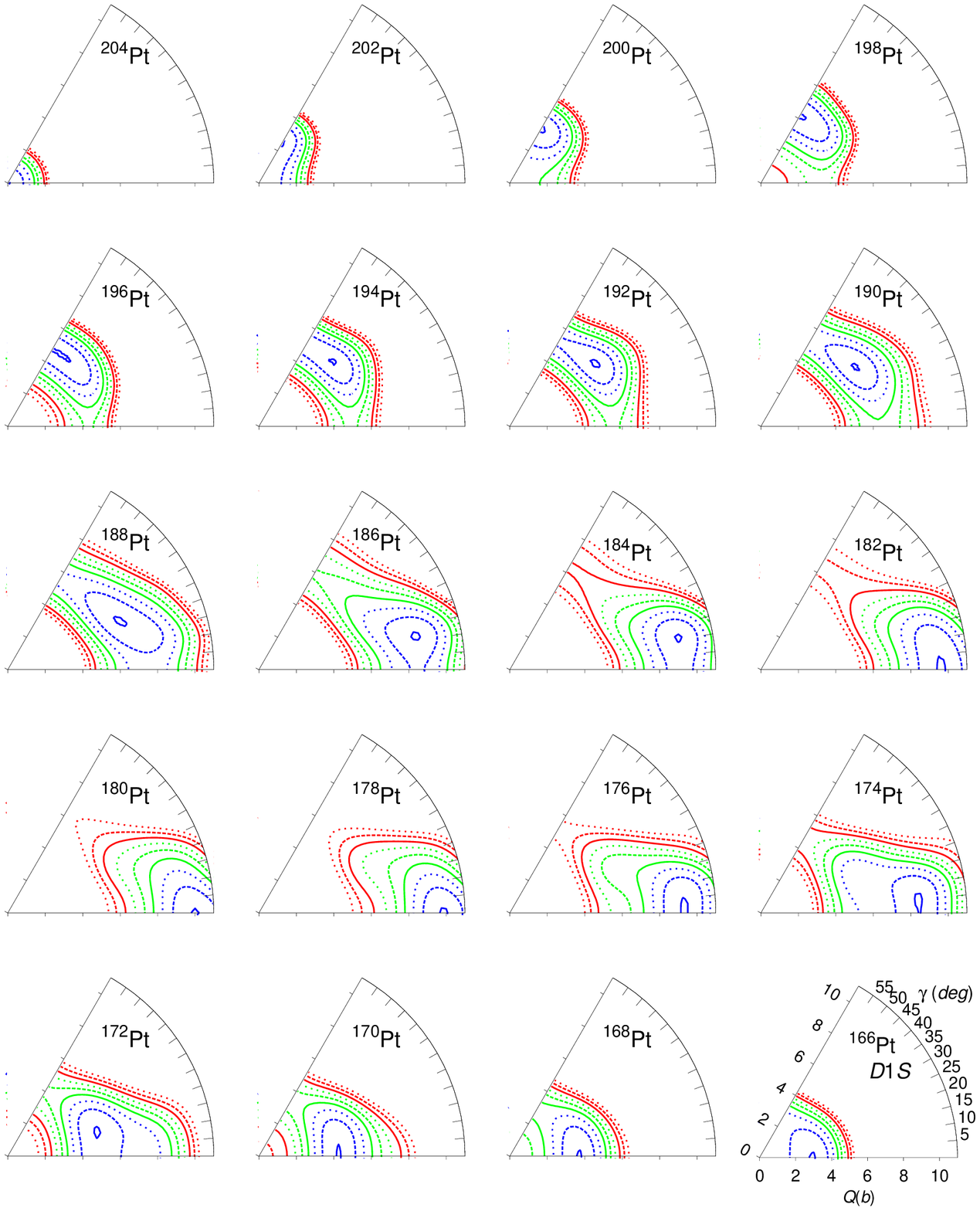}
\caption{(Color online) $Q- \gamma$ planes computed
with the Gogny-D1S force for the isotopes $^{166-204}$Pt.
The range of $Q$ values has been reduced to focus 
on the interval around the minima. The contour lines 
extend from the minimum up to 2 MeV in steps of 0.25 MeV.
Blue (black) contours are the three lowest, green (light gray) 
ones the next three and red (dark gray) contours
correspond to the three with higher energies. 
For each color the full line corresponds to the lower energy,
dashed to the next contour and dotted to the higher energy.
The minimum of the 
triaxial landscape can be identified by the small ellipse
surrounding it.
}
\label{fig_d1s}
\end{figure*}

\section{Theoretical Framework.}
\label{formalism}
In order to compute both PECs and PESs we have used the 
(constrained) HFB  method 
together with the parametrizations D1S, D1N
and D1M 
 of the 
Gogny interaction. The solution of the HFB 
equations, leading to the vacuum $| \Phi_{\rm HFB} \rangle$, was
 based on the so called gradient method
\cite{gradient,ours2} to locate the minima. The kinetic energy of the center of 
mass motion has been subtracted from the Routhian to be minimized in
 order to ensure 
that the center of mass is kept at rest. The exchange 
Coulomb energy was considered in the Slater approximation and
we neglected the contribution
of the Coulomb interaction to the pairing field.

The HFB  quasiparticle operators  have been expanded 
in an Harmonic Oscillator (HO) basis containing enough 
number of shells (i.e., N= 13 major shells) to grant 
convergence for all values of the mass quadrupole operator 
and for all the nuclei
studied. Energy contour plots will be shown in the $(Q,\gamma)$ 
plane \cite{ours2} (instead of $(\beta,\gamma)$)
with 

\begin{equation}
\label{Q20}
Q_{20} = \frac{1}{2} \langle \Phi_{\rm HFB} | 2z^{2} - 
x^{2} - y^{2}  | \Phi_{\rm HFB} \rangle 
\end{equation}

\begin{equation}
\label{Q22}
Q_{22} = \frac{\sqrt{3}}{2} \langle \Phi_{\rm HFB} |  x^{2} - y^{2}  | \Phi_{\rm HFB}
\rangle
\end{equation}

\begin{equation}
\label{Q0}
Q = \sqrt{  Q_{20}^{2}+ Q_{22}^{2}}
\end{equation}

\begin{equation}
\label{gamma}
\tan \gamma = \frac{Q_{22}}{Q_{20}}
\end{equation}

%%%%%%%%%%%%%%%%%%%%%%%%%%%%Fig3%%%%%%%%%%%%%%%%%%%%%%%%%%%%%%%%%%%%%%%%%%%%%%%%%%%%%%%%%
\begin{figure*}
\includegraphics[width=0.98\textwidth]{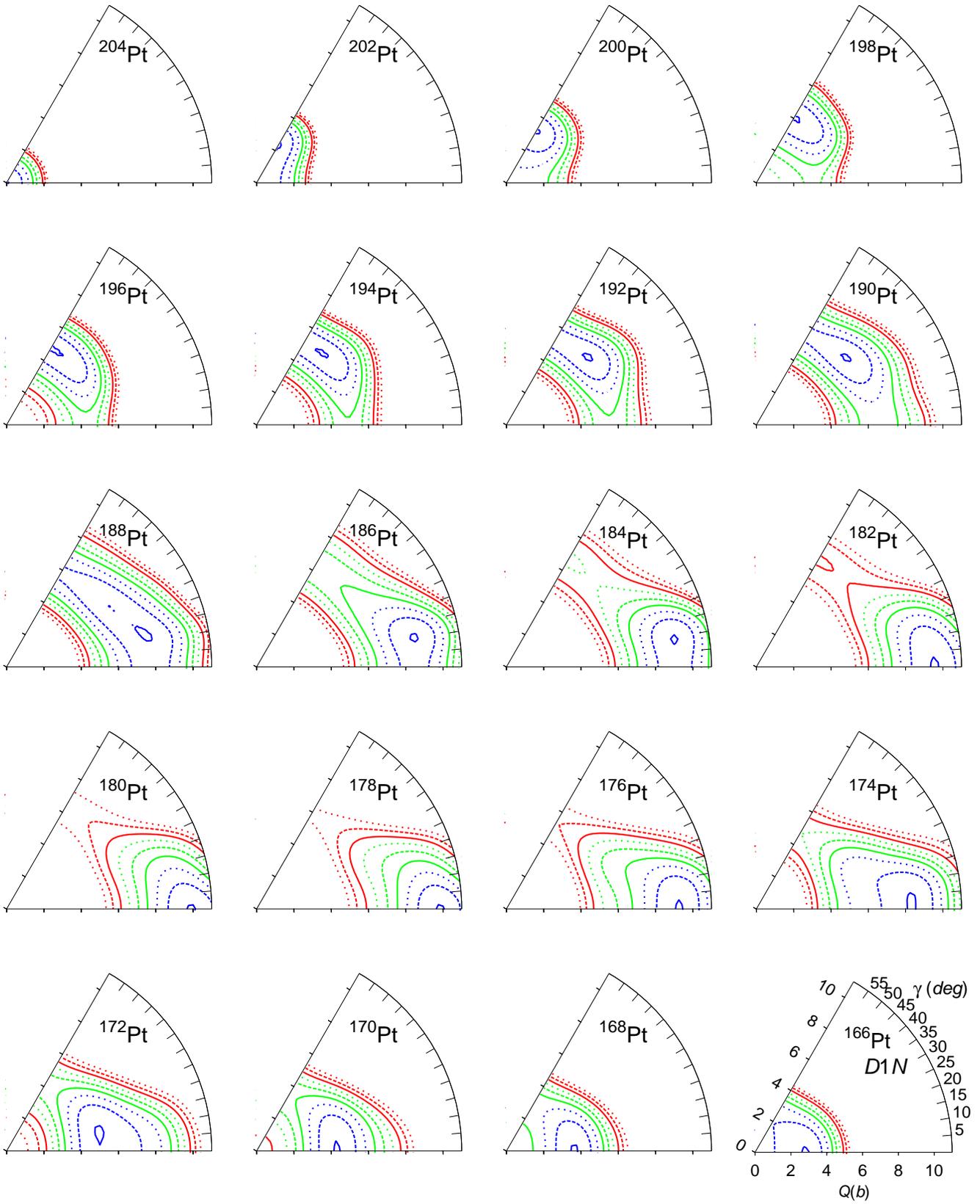}
\caption{(Color online) The same as Fig. \ref{fig_d1s}
but for the Gogny-D1N force.
}
\label{fig_d1n}
\end{figure*}

Other interesting pieces of information coming from the mean field
are the single particle energies (SPEs) for protons and neutrons. In our
calculations, with the Gogny interaction, we are solving the HFB equations
and therefore the only quantities that can be properly defined are the 
quasiparticle energies. However, in order to have  the more 
usual, Nilsson-like, diagrams  we have chosen to plot the eigenvalues
of the Routhian  \cite{rs}
$h = t+ \Gamma - {\lambda}_{20}Q_{20} - {\lambda}_{22}Q_{22}$, with 
$t$ being the kinetic energy operator, $\Gamma$ the Hartree-Fock
field. The term ${\lambda}_{20}Q_{20} +{\lambda}_{22}Q_{22}$
contains  the  
Lagrange multipliers used to enforce the corresponding 
constraints.
We have first performed calculations 
restricted to axially symmetric  shapes and in a second step
triaxiality is included in our mean field  analysis. In the first
case, obviously, the term  ${\lambda}_{22}Q_{22}$
is missing. In addition, the usual mean field constraints 
on both neutron and proton average numbers  are 
taken into account. Parity and time-reversal are selfconsistent
symmetries in our axial calculations whereas 
parity and simplex are the ones imposed in the 
triaxial case \cite{ours2}.

\section{Discussion of the results.}
\label{results}
In this section, we discuss the results of the present study. The 
systematics of deformation obtained for the isotopes $^{166-204}$Pt
is described in 
Sec. \ref{results-PECPES}. Single particle properties
are considered in Sec. \ref{results-SPE}.

\subsection{Mean field  systematics of deformation  for $^{166-204}$Pt.}
\label{results-PECPES}

The PECs obtained for the isotopes $^{166-204}$Pt  with our 
constrained  Gogny-HFB calculations
preserving axial symmetry, are shown 
in Fig. \ref{fig_axial} as functions of the  quadrupole
moment $Q_{20}$. Both prolate ($Q_{20} > 0$) and oblate
($Q_{20} < 0$) sides are displayed. The prolate side 
is equivalent to the triaxial results, to be discussed later on, 
with $Q=Q_{20}$ and $\gamma = 0^\circ$ whereas the oblate 
side is equivalent to $Q=|Q_{20}|$ and $\gamma = 60^\circ$.

  As we can see, the 
interactions  D1N and D1M provide PECs which are
extremely similar to the 
ones obtained with Gogny-D1S.
The deformations of the oblate and prolate minima are
practically independent of the force. 
The axial quadrupole
 moment $Q_{20}$ corresponding to the absolute minima of the 
 PECs increases  until $A\approx 180$ ($N\approx 102$) 
when it reaches the value $Q_{20} \approx 10$ b.
A sudden prolate to oblate shape change occurs around $A=188$ for
 all the Gogny
 interactions considered in the present study.
Beyond $A=190$, absolute oblate minima are obtained
with quadrupole moments decreasing until 
the spherical shape is reached for  $^{204}$Pt ($N=126$). The opposite 
situation occurs with the 
secondary minima. On the other hand, the depth of both 
prolate and oblate wells (as compared
to the spherical maximum) increases with increasing
 $N$ up to $A=182$ (roughly mid-shell) and it decreases from 
 there on being
the decrease more pronounced for the prolate wells. This, as explained 
in Ref. \cite{ours2} can be understood first as a consequence of 
the filling of down slopping levels coming from the high $j$, unique 
parity $i_{13/2}$ neutron orbital that
would explain the increase of the depth and then, at 
mid shell, the filling of the up slopping
levels that would lead to the decrease of the height of the wells.

%%%%%%%%%%%%%%%%%%%%%%%%%%%%Fig4%%%%%%%%%%%%%%%%%%%%%%%%%%%%%%%%%%%%%%%%%%%%%%%%%%%%%%%%%
\begin{figure*}
\includegraphics[width=0.98\textwidth]{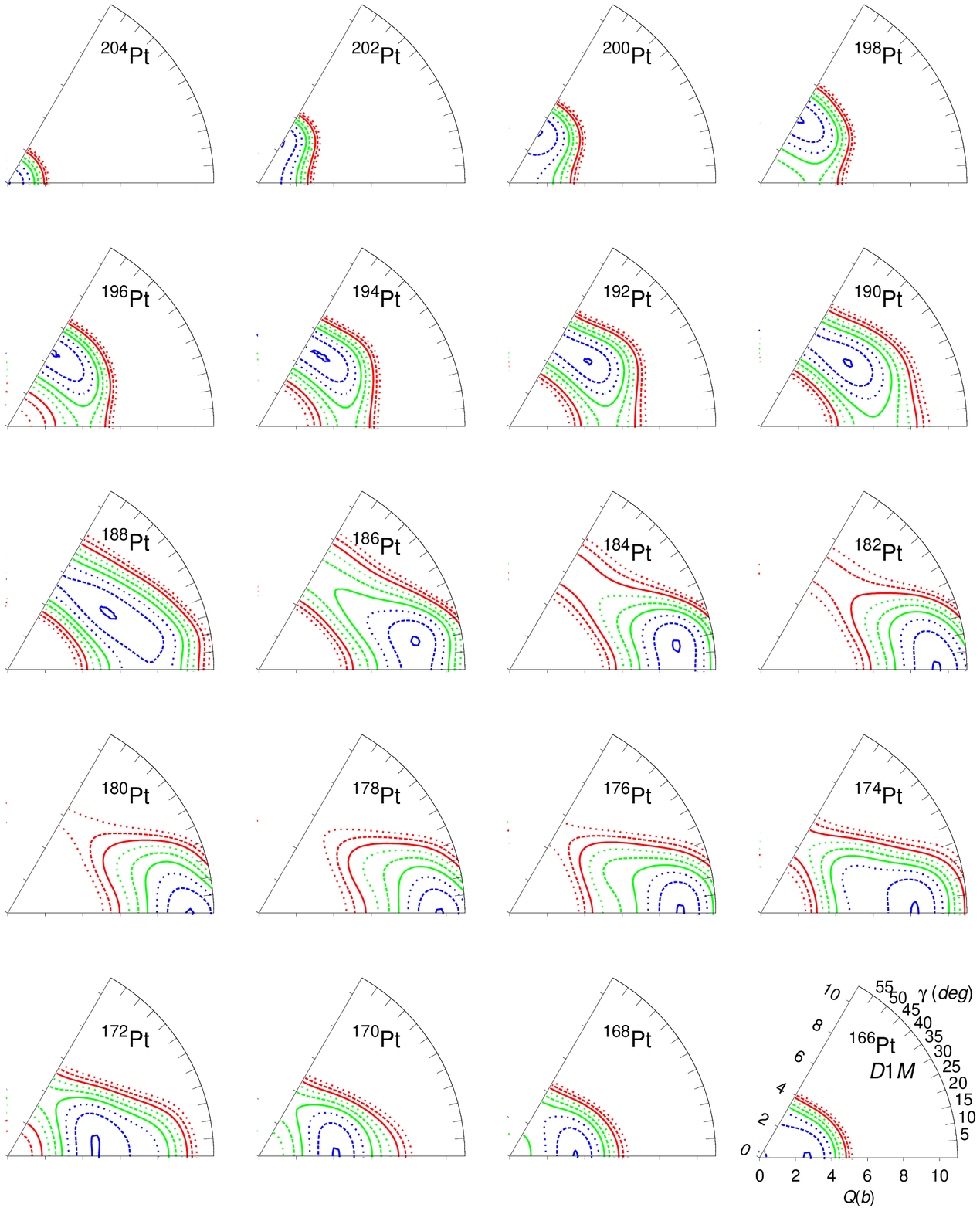}
\caption{(Color online) The same as Fig. \ref{fig_d1s}
but for the Gogny-D1M force.
}
\label{fig_d1m}
\end{figure*}

Slightly lower spherical barrier 
heights (i.e., the difference between the energy of the
absolute minimum of the PEC and the energy of the spherical configuration) are
predicted by the new Gogny forces D1N and D1M. Such a 
sensitivity of the spherical barriers with 
respect to  details of the effective interactions 
used has already
been found in previous studies \cite{Robledo-2,Rayner-1,Tajima,Sarri-Moreno}.
In our calculations, the largest and smallest  values of the
 total pairing energies corresponding 
to the spherical configurations are obtained with the sets
D1M and D1S, respectively. The
set D1N provides pairing energies in between.
This already reflects the different pairing content 
of the considered Gogny functionals but we postpone 
a discussion on this point for later on. 

%%%%%%%%%%%%%%%%%%%%%%%%%%%%Fig1%%%%%%%%%%%%%%%%%%%%%%%%%%%%%%%%%%%%%%%%%%%%%%%%%%%%%%%%%
\begin{figure*}
\includegraphics[width=0.89\textwidth]{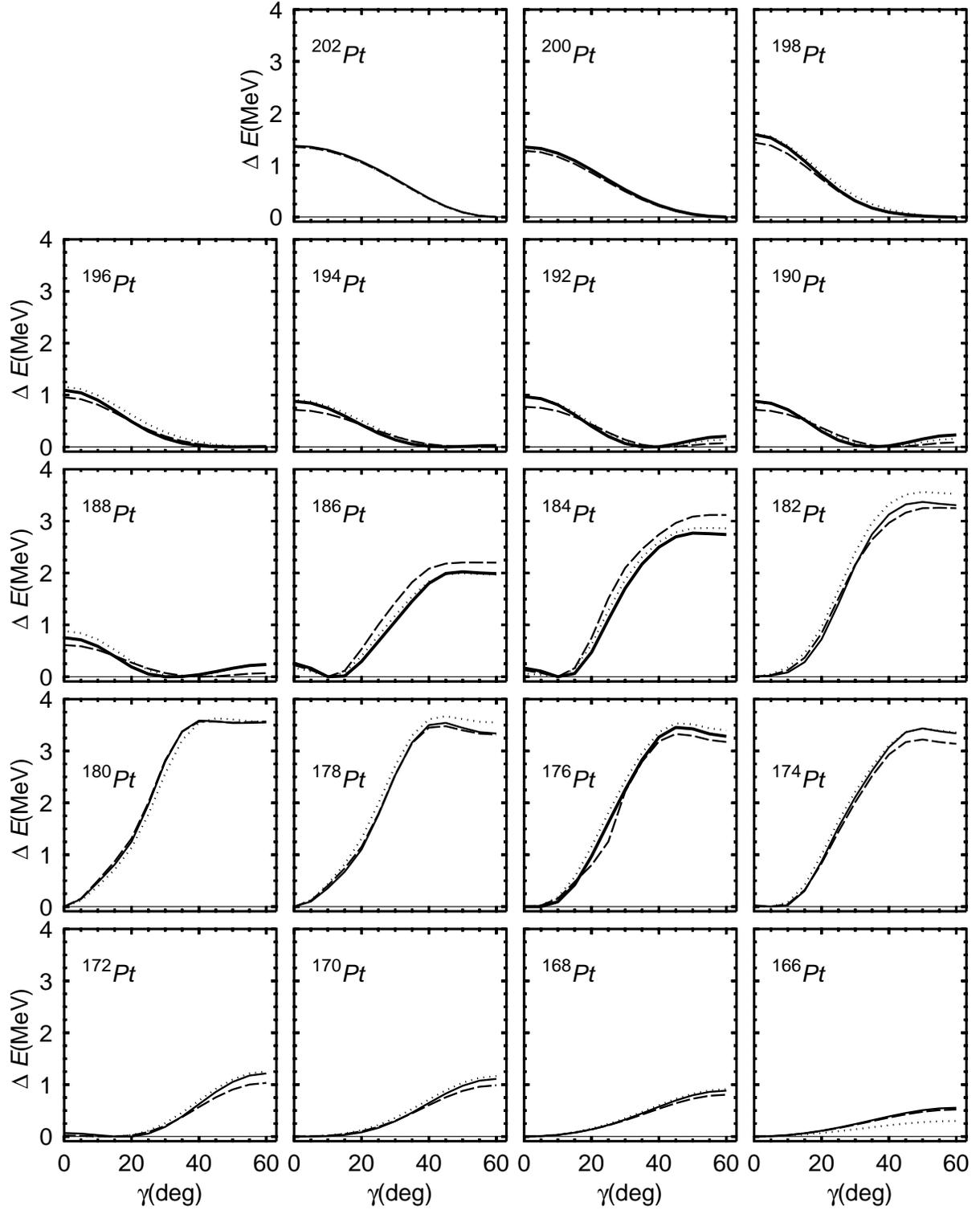}
\caption{Mean field excitation energies $\Delta E$ computed with the 
Gogny interaction D1S (continuous line), D1N (dashed line)
and D1M (dotted line) are
displayed as functions of the deformation parameter $\gamma$ for fixed
values of the quadrupole moment $Q$ corresponding 
to the lowest minima of the axially symmetric calculations 
(see Fig. \ref{fig_axial}).
Results for $^{204}$Pt  
are not included due to the presence of a spherical ground state.}
\label{evol-min-gamma}
\end{figure*}

The absolute values 
$| \Delta E_{p-o} |$
of the energy differences 
between the prolate and oblate
minima of the PECs  exhibit two bumps, the
first with a maximum  at $A=178$ 
(2.19, 1.97 and 2.11 MeV for D1S, D1N and 
D1M) corresponds to
isotopes with a prolate ground state
and the second one at  $A=198$
(1.16, 0.95 and 0.99 MeV for D1S, D1N and 
D1M) corresponds to isotopes
with an oblate ground state. The nucleus $^{188}$Pt, separating
 the two regions, 
has almost degenerate 
prolate and oblate minima with 
$| \Delta E_{p-o} |$ values of 
45, 111 and 71 keV for  D1S, D1N and 
D1M, respectively. 
  
\begin{figure*}
\includegraphics[width=0.98\textwidth]{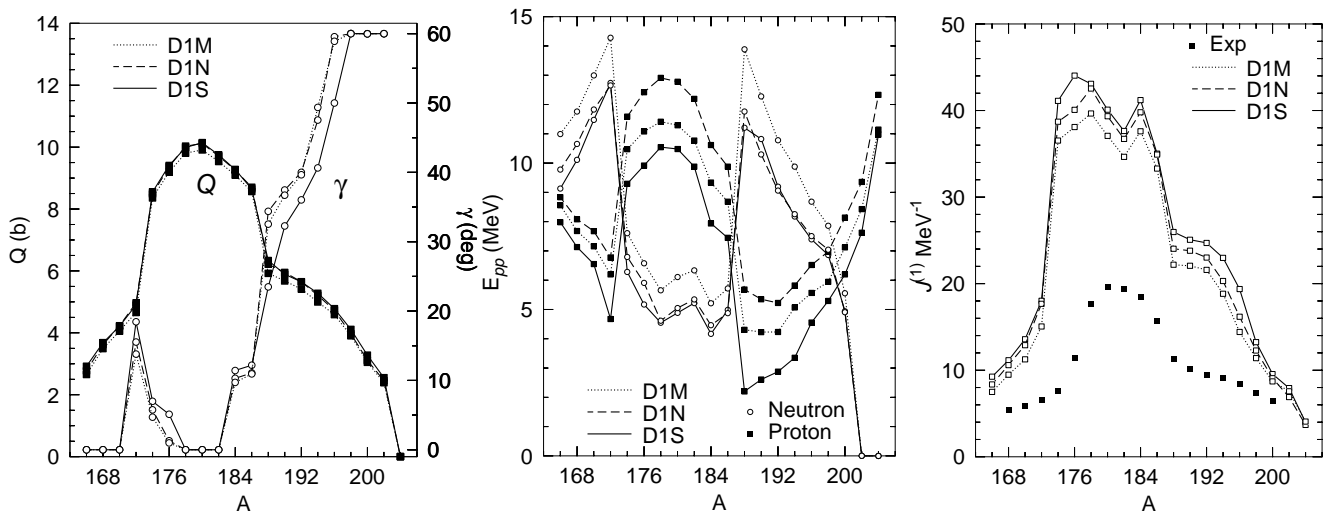}
\caption{Ground state triaxial coordinates 
($Q,\gamma$) (left panel), neutron and proton 
pairing energies (middle panel) and Thouless-Valatin
moments of inertia (right panel) are plotted 
as functions of the mass number A. Calculations have 
been performed with the Gogny interactions D1S, D1N and 
D1M. 
}
\label{figevolQ}
\end{figure*}

The previous axially symmetric results, agree well with the ones 
in Ref. \cite{Robledo-2} using the parametrization SLy4 \cite{sly4}
of the Skyrme interaction in the particle-hole channel plus a 
zero range and density dependent pairing  
interaction \cite{Terasaki} (with strength 
$g= 1000$ MeV fm$^{3}$ for both protons and neutrons)
and with previous Skyrme-HF+BCS calculations with the parameter set 
SIII \cite{Sauvage}.
They also agree well with the results of the axial 
calculations reported in Ref. \cite{Sharma-Ring} using the parametrizations
NL1 and NL2 of the relativistic mean field Lagrangian and 
with axial macroscopic-microscopic calculations reported in Ref. \cite{Moller-1995}.
 However, in our axial calculations, 
 prolate and oblate minima lie 
quite close in energy ($| \Delta E_{p-o} | \le 2.2 $ MeV).
Thus, a 
$\gamma$-path connecting them would be possible
and triaxiality could play a role, as will be discussed 
below, converting some of the axially symmetric minima 
into saddle points. Therefore, in a second step, we have also explored 
the triaxial landscape and construct PESs
for all the considered nuclei.

The PESs obtained with the Gogny sets D1S, D1N and 
D1M 
are shown in Figs. \ref{fig_d1s}, \ref{fig_d1n} and \ref{fig_d1m} 
in the form of $Q-\gamma$ planes.
In order to simplify the presentation, the range
of Q values plotted  
is reduced to 
0 b $\le Q \le 11$ b and the contour lines are
also severely  reduced by considering contours every 
250 keV  up to 2 MeV higher than the energy
of the minimum. We can see, the spherical structure 
in $^{204}$Pt while  $^{202-198}$Pt
exhibit  oblate 
($\gamma =60^\circ$)
ground states with $Q \approx$ 3-4.5 b. An 
{\it{island of triaxiality}}, centered around the nucleus 
$^{188}$Pt,
is clearly visible from these figures. The ground 
state triaxial coordinates
$(Q$,$\gamma)$ within such an {\it{island of triaxiality}}
evolve  from
$(Q \approx $5 b, $\gamma \approx 50^\circ-58^\circ)$ in $^{196}$Pt
to $(Q \approx $9 b, $\gamma \approx 10^\circ)$ in 
$^{184}$Pt. The depth of these triaxial
ground states, compared with the axial ones,  is very small
(see below). In the case of $^{182-178}$Pt, our 
calculations predict
prolate ($\gamma = 0^\circ)$
ground states  with 
deformations  $Q \approx 10 $ b. 
Again, for $^{176-172}$Pt  
 very shallow
triaxial minima are predicted. In the case of 
$^{172}$Pt, for example, we find
$(Q \approx $5 b, $\gamma \approx 20^\circ)$.
The isotopes $^{170-166}$Pt display  prolate ground states 
with $Q \approx 3-4.5$ b.

%%%%%%%%%%%%%%%%%%%%%%%%%%%%Fig7%%%%%%%%%%%%%%%%%%%%%%%%%%%%%%%%%%%%%%%%%%%%%%%%%%%%%%%%%
\begin{figure*}
\includegraphics[width=0.98\textwidth]{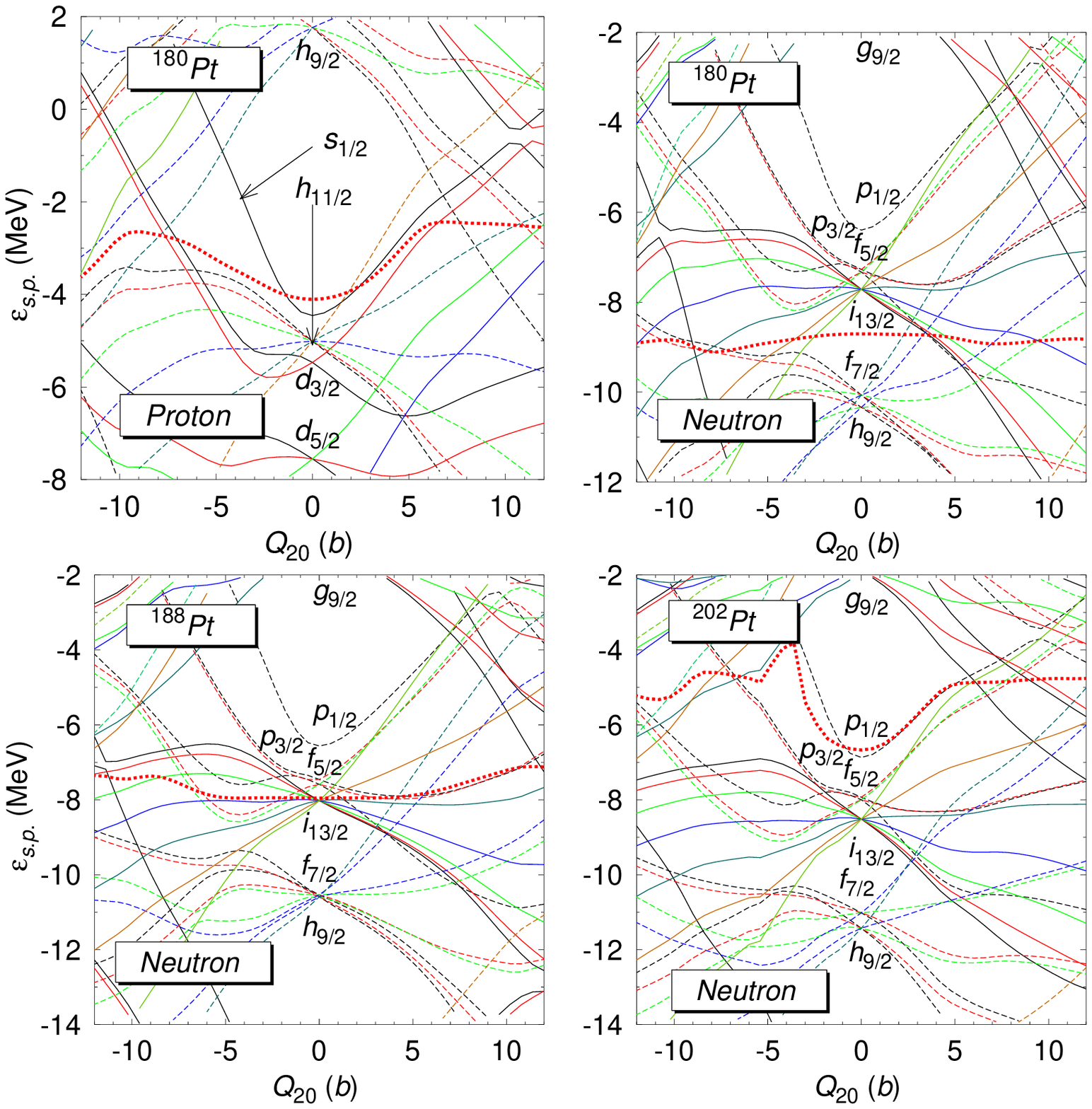}
\caption{(Color online) Upper panels: Proton (left panel)
and neutron (right panel) SPEs
for the nucleus  $^{180}$Pt
as  functions of the axial quadrupole moment $Q_{20}$.
The Fermi levels
are also plotted with a thick (red) dotted line.
The results correspond to the force Gogny D1S. 
Solid (dashed) lines are used for positive (negative) parity states.
With
increasing $K=1/2,3/2,5/2, \dots$ values color labels 
are black, red, green, blue, dark-blue, brown, dark-green, etc.
Lower panel: the same as above but for neutron SPEs 
of the nuclei $^{188}$Pt (left panel) and 
$^{202}$Pt (right panel). The spherical quantum numbers at $Q_{20}=0$
are given for a number of orbitals close to the Fermi level.
}
\label{speq20}
\end{figure*}

%%%%%%%%%%%%%%%%%%%%%%%%%%%%Fig8%%%%%%%%%%%%%%%%%%%%%%%%%%%%%%%%%%%%%%%%%%%%%%%%%%%%%%%%%
\begin{figure*}
\includegraphics[width=0.95\textwidth]{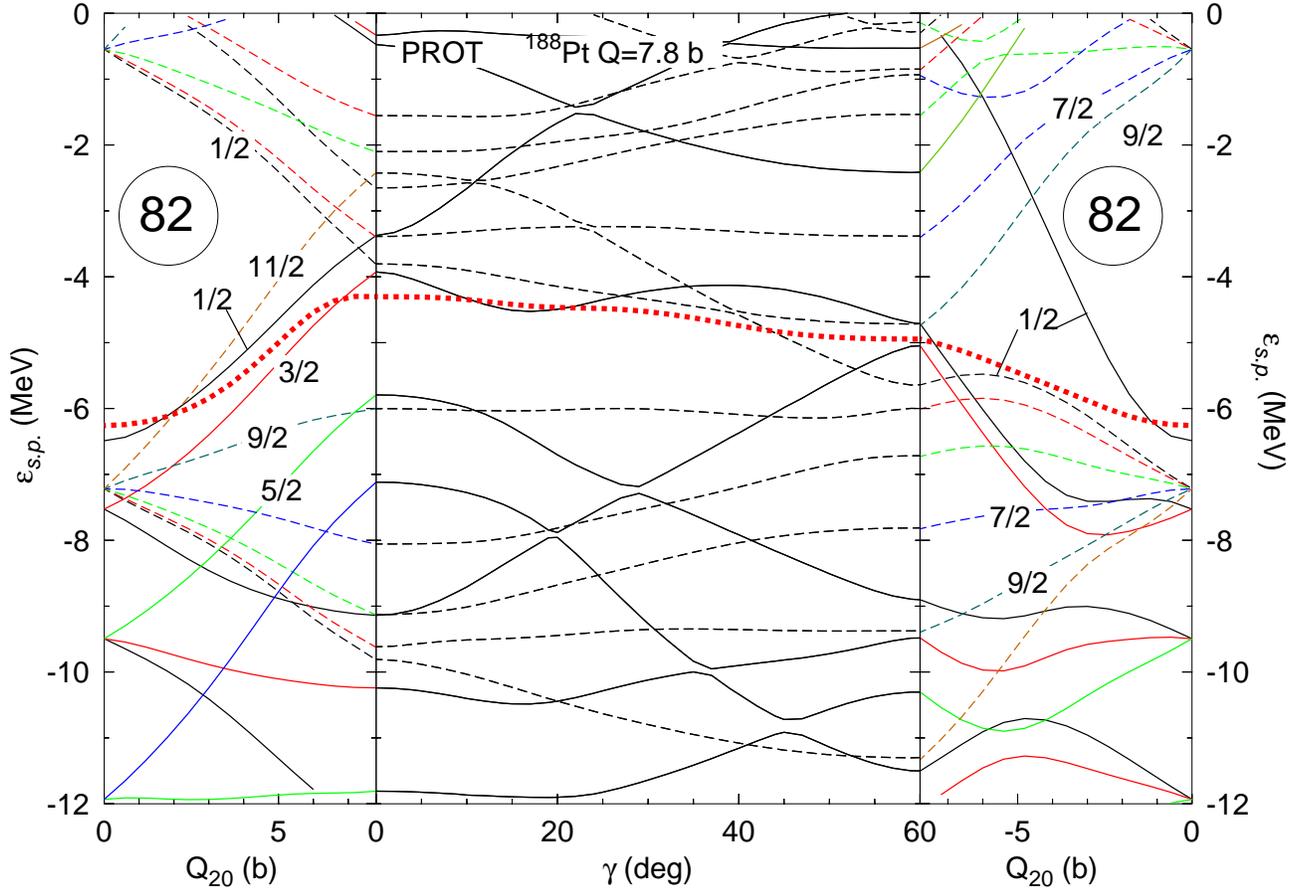} 
\caption{(Color online) Proton 
SPEs are plotted for the nucleus $^{188}$Pt. In the left panel, the SPEs are plotted
as a function of the axially symmetric quadrupole moment $Q_{20}$
from  $Q_{20}=0$ up to $Q_{20}=7.8$ b. In the middle part the triaxial
proton SPEs are plotted as functions of the  deformation
parameter  $\gamma$ and for $Q=7.8$ b. Finally, on the right panel, the axially symmetric
SPEs are plotted
as a function of $Q_{20}$ from  $Q_{20}=-7.8$ b up to $Q_{20}=0$. The proton
Fermi level is depicted as a thick (red) dotted line. Some relevant $K$ values
are also included. In the axial plots, at $Q_{20}=0$ (i.e. sphericity) we
have  the $d_{5/2}$   at $\epsilon \approx -9.5$ MeV, at an energy of
around -7.5 MeV we have the $d_{3/2}$  and at $\epsilon \approx -7.2$ MeV we have 
the $h_{11/2}$ orbital. Also at $Q_{20}=0$ the $Z=82$
shell gap is clearly visible. 
Solid (dashed) lines are used for 
positive (negative) parity states. For more details, see the main text.
}
\label{speAXTRI-proton}
\end{figure*}

In order to obtain a more quantitative understanding of the 
PESs we have plotted in Fig. \ref{evol-min-gamma}, the mean field
energies corresponding 
to the lowest  axial minima $Q=Q_{20}$
of the PECs  in Fig. \ref{fig_axial}
as functions of the deformation parameter $\gamma$. We
 observe that only one
  of the two axial minima 
remains in most of the cases. The nuclei 
$^{202-198}$Pt exhibit  oblate $(\gamma =60^\circ)$ 
absolute minima and the
 prolate (axial) solutions become saddle points with excitation
  energies $\Delta E \le 2 $ MeV. On the other hand, $^{196-188}$Pt are 
  rather $\gamma$-soft with triaxial minima  almost
degenerate with the (axially symmetric) prolate and oblate solutions. 
In the case of $^{188}$Pt, for example, we 
obtain $|\Delta E|_{triaxial-saddle} \le 0.9 $ MeV.
Still inside the {\it{island of triaxiality}}, the oblate 
configurations  in both $^{186,184}$Pt already show the tendency
to increase their excitation energies. A similar  trend 
for the oblate solutions
is 
observed within the mass range $182 \le A \le 174$.  
Oblate and prolate configurations for the nuclei 
$^{172-166}$Pt are quite close
($0.3 \le \Delta E \le 1.2$ MeV) and 
softly linked  along the $\gamma$ direction. 

A  detailed account of the evolution of
the ground 
state triaxial coordinates $(Q,\gamma)$ as functions of the mass 
number $A$ is  presented in the left  panel of 
Fig. \ref{figevolQ} for the sake of completeness. The 
striking similarity of the ground state deformations
obtained with   
the Gogny interactions D1S, D1N and D1M becomes evident from this plot. 
We observe the emergence of weakly oblate  ground 
states for the isotopes $^{202-198}$Pt. On the other hand, the 
sudden shape
transition observed in the framework of the axially symmetric
HFB calculations is now
replaced by a smooth shape change through
the {\it{island of triaxiality}} 
represented, in our case, by the isotopes
$^{184-196}$Pt. A prolate
deformed regime is predicted for the isotopes 
$^{178-182}$Pt. We find 
that, the  trend
of shape changes predicted by our calculations
in the considered Pt isotopes
agrees well with the  ones 
obtained in 
Refs. \cite{Ansari,Krieger} 
and the conclusions
extracted from the combination of total routhian surface 
calculations plus  quasiparticle random phase approximation
(TRS+QRPA) and IBM models in  Ref. \cite{Cederwall-exp}.
The general trend in our calculations is also
consistent with results obtained in the framework of the 
Strutinsky approach \cite{bengtsson,Moller-2008}. 
Further down in neutron number, the TRS results 
\cite{review,Ce-other}
predict a rapid  change to a  triaxial shape
also found in our calculations around $^{172}$Pt.
Finally, our PESs  for  $^{170-166}$Pt
exhibit features that could 
be interpreted as the onset of a more pronounced 
vibrational character for these nuclei \cite{Cederwall-exp}.

%%%%%%%%%%%%%%%%%%%%%%%%%%%%Fig9%%%%%%%%%%%%%%%%%%%%%%%%%%%%%%%%%%%%%%%%%%%%%%%%%%%%%%%%%
\begin{figure*}
\includegraphics[width=0.95\textwidth]{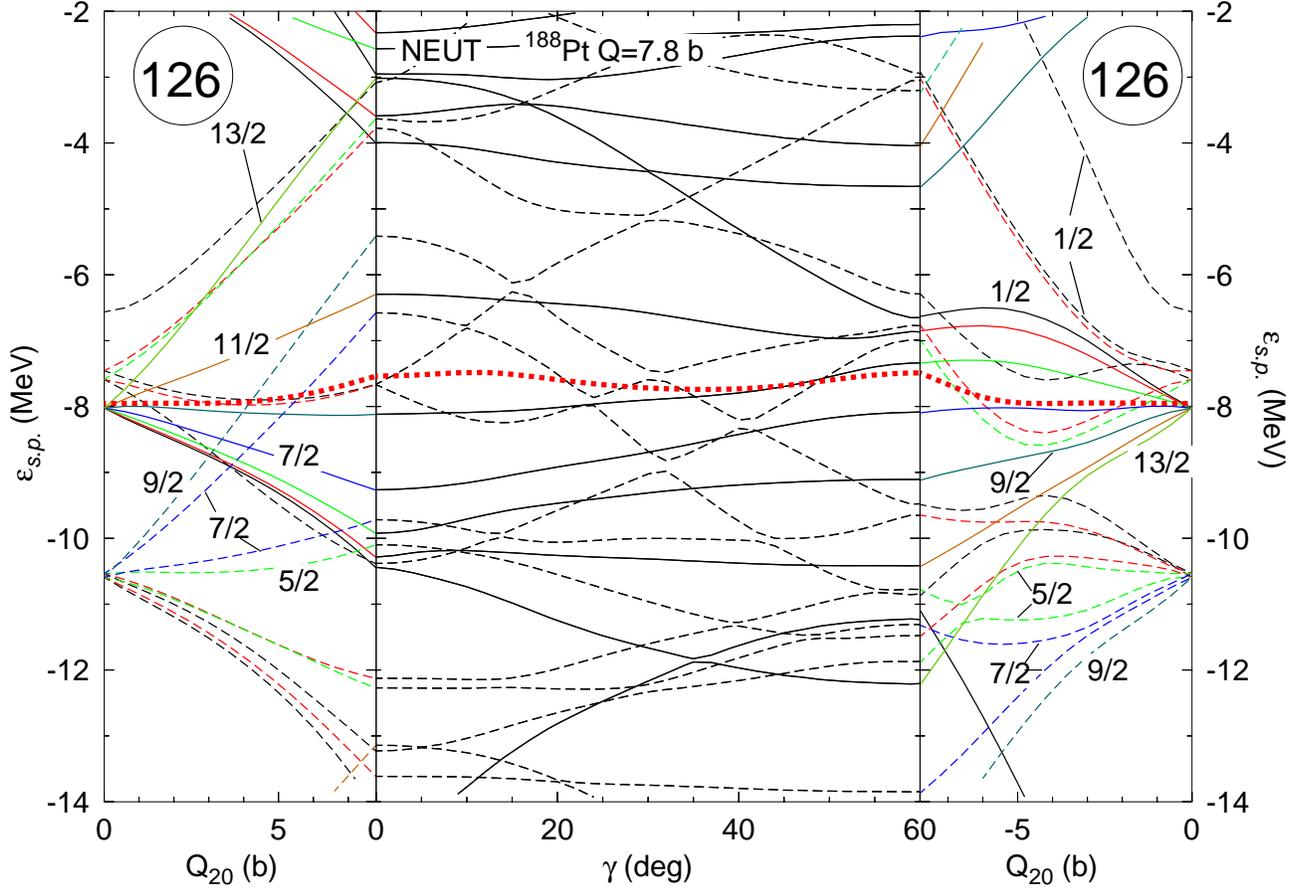} 
\caption{(Color online) The same as Fig. \ref{speAXTRI-proton} but 
for neutrons. In the axial plots, at $Q_{20}=0$ (i.e. sphericity) we
have  the $h_{9/2}$ and $f_{7/2}$  at $\epsilon \approx -10.5$ MeV, at an energy of
around -8 MeV we have the $i_{13/2}$  and at $\epsilon \approx -7.5$ MeV we have the 
$f_{5/2}$ and $p_{3/2}$ orbitals. Also at $Q_{20}=0$ the $N=126$
shell gap is clearly visible. 
}
\label{speAXTRI-neutron}
\end{figure*}

The rather involved behavior of the neutron and 
proton pairing 
energies  $E_{pp}$ (with opposite sign) 
versus mass number 
for the Pt isotopes is shown 
in the middle panel
of Fig. \ref{figevolQ}. As expected for a pure $N=126$ 
shell closure, neutron pairing collapses for the 
nucleus $^{204}$Pt. We
observe that proton pairing shows a non constant behavior 
in spite of having constant proton number ($Z=78$)
that comes from selfconsistency effects. On the other hand, neutron 
pairing energies are lower inside 
the region between $A=174$ and $A=186$ that is precisely where the 
strong prolate deformation develops.
The lowering of pairing energies is a consequence of the lowering of the level density that is
needed (Jahn-Teller effect) to induce the deformed minima. For other 
values of $A$, the neutron level
density around the Fermi level is higher and, as a consequence, pairing
 correlations are stronger.
Concerning different values of the neutron and proton pairing, 
for the three Gogny functionals  the pairing energies follow
 the same isotopic trend 
and the only relevant change is in the absolute value 
that tends to be slightly lower for D1S. At this
point it is worth to remember that the value of 
the pairing  energy shown is
related to the amount of pairing correlations 
present in the system but it is  by far not certain that
the correlation is linear, in other words, the 
different values of $E_{pp}$ for different interactions
do not necessarily imply the same quantitative 
behavior for pairing correlations.

Finally, the Thouless-Valatin moments of inertia 
$J^{(1)}= 3/\left(E_{2^{+}}- E_{0^{+}}\right)$
of the first 2$^{+}$ states 
in $^{166-204}$Pt are 
plotted in the right panel  of Fig. \ref{figevolQ} 
as functions of the mass number
$A$. They are compared with the experimental values
extracted 
from the available systematics for the excitation energies 
of the 
first 2$^{+}$ states 
in $^{166-204}$Pt (see, for example, \cite{Cederwall-exp}).
The energies needed for the computation of 
$J^{(1)}$ have been obtained using the 
selfconsistent cranking approximation introducing 
the usual  time reversal breaking  
term $-\omega J_{x}$ and the subsidiary 
condition $\langle  J_{x}  \rangle= \sqrt{I(I+1)}$
\cite{rs,ours2,gogny-other-6}. The Thouless-Valatin 
moments of inertia strongly depend on pairing 
and therefore a comparison of the results obtained 
with the three  Gogny functionals considered in this
work, which exhibit different pairing contents, can also give  
a hint on the quality of their predictions. As can be seen, the results 
follow the same isotopic trend irrespective of the Gogny force 
with the tendency to be the largest 
for Gogny-D1S and the smallest in the case 
of Gogny-D1M. Nevertheless, the differences in the predicted 
values can still be attached to the uncertainties in the 
effective interactions and we observe how the
selfconsistent cranking results tend to overestimate 
the experimental values.
This defect, well known already, is a direct consequence of too
low pairing at the mean field level and its solution would
require an improved treatment of pairing correlations. There
are many mechanisms beyond mean field that modify the amount
of pairing correlations in a given nuclear system, but there are two
particularly important that tend to increse correlations. One is
the restoration of the particle number symmetry broken by the
HFB method and the other is shape fluctuations around the HFB minimum.
The latter is connected with the fact that the HFB minimum corresponds to
a low level density region of the single particle spectrum (see next
subsection)
and therefore its amount of pairing correlations is far lower than
the one of the neighbouring configurations. Taking into account
the first mechanism would involve particle number projection, whereas
the second can only be treated in the scope of the GCM with
the quadrupole moment as generating coordinate or in the Bohr Hamiltonian
method \cite{delaroche10}. Clearly, both methods are out of the scope of
the present study.
On the other hand, we observe certain correlation between
the evolution  with the number of neutrons of the quadrupole 
moments $Q$ and the moments of inertia $J^{(1)}$, as it is apparent
by looking at the left and right panels of Fig. \ref{figevolQ}.
This correlation is not so evident for the $\gamma$ degree of freedom.

The results discussed in this Section indicate, that the new 
interactions/functionals D1N  and 
D1M  provide
the same quality of mean field  ground state
predictions for the considered Pt isotopes as compared 
with the  Gogny-D1S  
force taken as a reference in our calculations.
This coincidence give us confidence
on the robustness of our  mean field 
predictions with respect 
to a change in the particular parametrization 
of the Gogny interaction used. The 
agreement is also rather good with the
mean field picture obtained in Refs. \cite{Robledo-2,ours2} within the HF+BCS framework 
based on the Skyrme parametrization SLy4 
in the particle hole 
channel plus a zero range and density dependent pairing 
(with strength 
$g= 1000$ MeV fm$^{3}$ for both protons and neutrons). Both
Gogny-D1S and Skyrme-SLy4 represent well reputed interactions
whose reasonable predictive power has already been 
thoroughly tested all over the nuclear chart 
and 
it is  very satisfying  to observe how the
new parametrizations D1N and D1M, in spite
of the relaxation of some of the original 
constraints in their fitting protocols and more 
oriented to reproducing nuclear
masses \cite{gogny-d1n,gogny-d1s}, still follow 
very closely the fine details predicted 
with Gogny-D1S, Skyrme-SLy4  and 
other theoretical models 
\cite{Robledo-2,ours2,bengtsson,Krieger,Ansari,Sharma-Ring,Sauvage,Cederwall-exp,Ce-other} 
for an isotopic chain
with such a challenging shape evolution.

Let us remark that we are perfectly aware of the fact that 
the (static) mean field picture 
described above, should also be 
extended to a dynamical treatment of the relevant 
degrees of freedom. This becomes clear from the topology
of the PESs  indicating that, in order
to access a  quantitative comparison of the energy 
spectra and reduced transition probabilities with the considered 
Gogny interactions, the dynamical interplay between the
zero point motion associated with the restoration 
of  broken symmetries (mainly, angular momentum
and particle number) and fluctuations in the collective
parameters $(\beta,\gamma)$ should be taken into account. For such 
a cumbersome and computer power demanding extension the 
Gaussian Overlap Approximation (GOA) appears as a first  
suitable choice \cite{bertsch,Libert-1,GOA-other-1}.
For a very recent and excellent pedagogical review, the 
reader is also referred to \cite{Prochniak}. Work
along these lines is in progress and will be reported 
elsewhere.

%%%%%%%%%%%%%%%%%%%%%%%%%%%%Fig10%%%%%%%%%%%%%%%%%%%%%%%%%%%%%%%%%%%%%%%%%%%%%%%%%%%%%%%%%
\begin{figure*}
\includegraphics[width=0.98\textwidth]{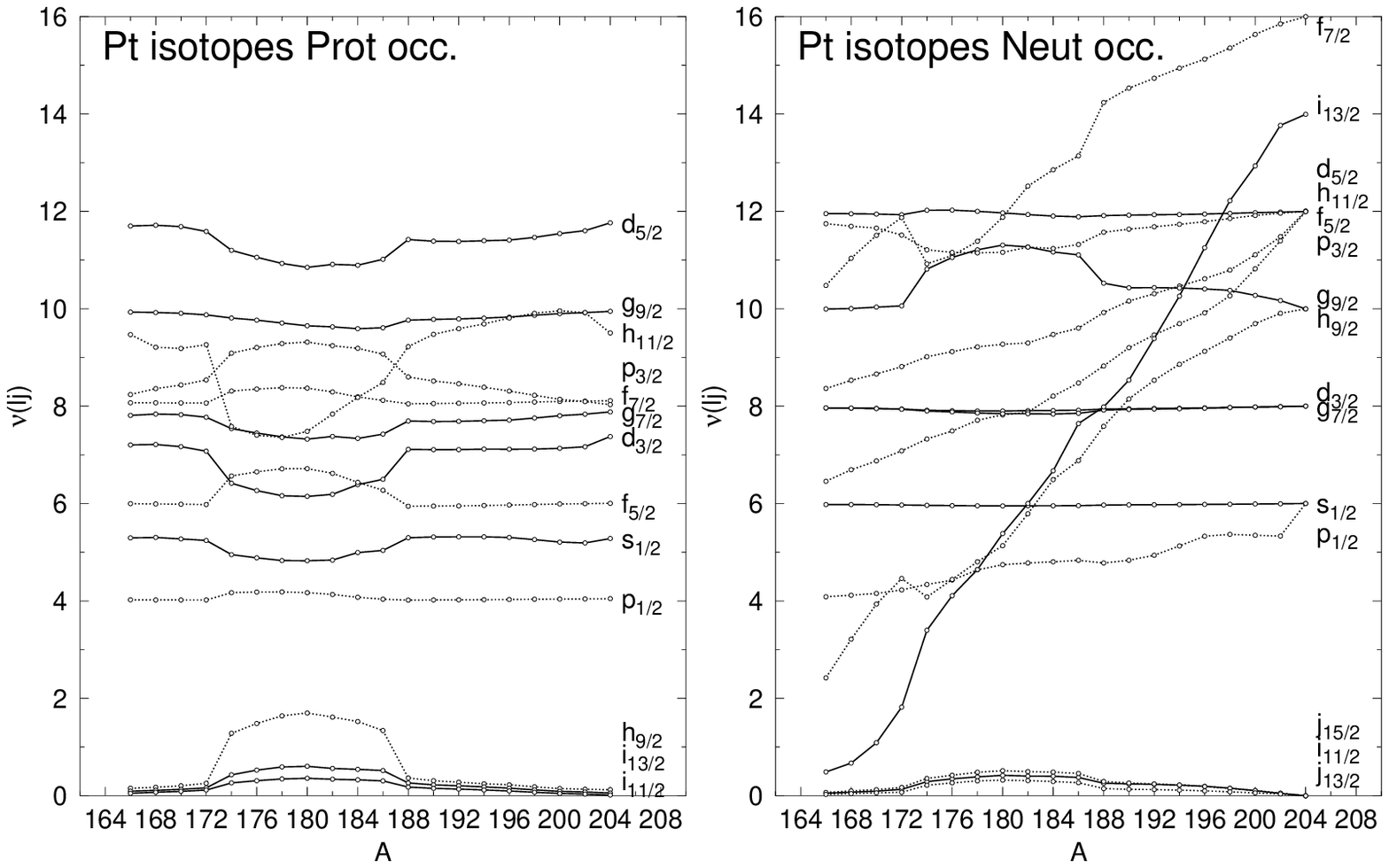}
\caption{Spherical occupancies in the 
proton (left panel) and neutron (right panel) 
ground state wave functions for the isotopes 
$^{166-204}$Pt. For details, see the main text.
}
\label{occupations}
\end{figure*}

\subsection{Single particle properties}
\label{results-SPE}
In this section, we pay  attention to single particle 
properties of the considered isotopes. To this end, we first
show in Fig. \ref{speq20}    
SPE 
plots, as functions of the 
axial quadrupole moment
$Q_{20}$. The isotopes $^{180}$Pt, $^{188}$Pt and 
$^{202}$Pt  are taken as illustrative examples. The SPEs 
correspond to our Gogny-D1S calculations. For
other nuclei
and/or  Gogny interactions  the results 
are  quite similar.

The  energy levels in Fig. \ref{speq20} 
gather together around the spherical configuration $Q_{20}$=0
forming the spherical shell model orbitals $nlj$. Due to axial
and time-reversal symmetries SPE levels, tagged by the 
$K$ quantum number corresponding to the third component of
the angular momentum in the intrinsic frame, are doubly
degenerate. Positive and negative parity states 
are plotted with full and dashed lines, respectively.
The proton ${\lambda}_{Z}$ and neutron ${\lambda}_{N}$
Fermi levels  
are also shown   
with a thick (red) dotted line.
As it is well known, atomic nuclei "avoid" regions 
with high single particle level densities (Jahn-Teller effect)
and therefore the plots of SPEs versus deformation
help us to identify regions where  
energy gaps favor the
appearance of deformed minima 
\cite{HW}.

We also look at  the onset of deformation in the 
considered Pt nuclei using the Federman-Pittel (FP)
criteria \cite{pittel}. The origin of nuclear deformation
is certainly a much more complicated phenomenon involving 
different mechanisms (see, for 
example, \cite{Werner,rayner-PRL,Naza-def,Mottelson}
and references therein for  detailed discussions 
on this issue). Even though the FP argument 
may be
 incomplete, it is 
certainly playing a role in the onset of nuclear deformation
and we resort to it in the present 
mean field study
as a way to establish a very qualitative and 
simplified overall picture of 
shape changes in terms of the evolution of the underlying 
SPEs. For recent studies using
similar ideas, see \cite{ours2,Fossion-def}. In order
to incorporate the terminology of spherical orbitals
to discuss deformed configurations, we assign to a given
deformed single particle orbital the label of the 
spherical orbit from which it originates at $Q_{20}=0$
\cite{ours2}.

In  Fig. \ref{speq20}, we can see at zero
deformation the spherical proton shells 
$3s_{1/2}$, $1h_{11/2}$, $2d_{3/2}$
and $2d_{5/2}$ below the Fermi level ${\lambda}_{Z}$  and the 
$1h_{9/2}$ level above it.
The relative position of 
${\lambda}_{Z}$, with respect to the SPEs, is 
quite stable. The main effect of the different neutron numbers 
appears in the scale of  Fermi energies ${\lambda}_{Z}$
ranging from around -9 MeV
in the case of  $^{204}$Pt  to values close
to zero in the case of the  neutron deficient isotope $^{166}$Pt.
Therefore, in Fig. \ref{speq20} we have only plotted the 
proton SPEs for the isotope $^{180}$Pt. 
The 
closeness to the proton magic number $Z=82$ makes 
Pt isotopes ($Z=78$) to display a tendency to be spherical or 
slightly oblate
as it can be seen in Fig. \ref{speq20}
from the huge spherical gap 
of $\sim$ 6 
MeV
and the large gap 
of  $\sim$ 3 
MeV
on the oblate side centered around $Q_{20}=-5$ b.

For neutrons the relevant  spherical orbitals
shown in Fig. \ref{speq20} are  
$1h_{9/2}$, $2f_{7/2}$, $1i_{13/2}$, $2f_{5/2}$,
$3p_{3/2}$, $3p_{1/2}$
and $1g_{9/2}$. Notice also that in the case  
of  $^{204}$Pt  
the tendency to be spherical is reinforced by its magic neutron
 number $N=126$ and the final result is a spherical nucleus.
In our calculations, the $N=126$ spherical shell gap
 is observed to change with mass number
from $\sim$  5.5 MeV in $^{180}$Pt to $\sim$ 4.5 MeV in $^{202}$Pt
going through $\sim$ 5 MeV for $^{188}$Pt.

Between  $^{202}$Pt
and $^{196}$Pt, ${\lambda}_{N}$ 
crosses a region on the 
oblate side where an energy gap of $\sim$ 3 MeV (i.e. half the size
of the spherical gap) is found. This occurs at
around  $Q_{20}=-5$ b, which overlaps perfectly with the gap
on the oblate proton sector already mentioned above.
On the other
hand, as can be seen from Fig. \ref{speq20}, the
$1i_{13/2}$ gets more and more occupied for increasing
neutron number $N$ and at a certain point (i.e., beyond $^{194}$Pt)
its role is transferred to the $2f_{5/2}$ and 
$3p_{3/2}$ orbitals. According to FP only the neutron
$2f_{5/2}$ can interact with the proton 
$2d_{3/2}$ 
(i.e., $n_{p}=n_{n}$, $l_{p}=l_{n}-1$) 
but since the $l$ values are in this case
low we should not expect  a very strong interaction among them.
This, qualitatively 
explains the appearance of the, very soft and close
to spherical, oblate minima displayed 
in Fig. \ref{fig_axial}
within the mass range $196 \le A \le 202$. 
A similar mechanism leads 
to very shallow and weakly prolate 
secondary minima in $^{196,198}$Pt.

Between $^{194}$Pt and $^{184}$Pt the neutron Fermi level 
on the 
prolate sector 
 crosses a region with energy gaps of $\sim$ 3 MeV and between
 $Q_{20}=3$ b and 10 b.
From the neutron SPE plot of $^{188}$Pt, we observe 
how within this mass 
range, the most  prominent role
is played by the   $1i_{13/2}$ 
which, according to the FP criteria, interacts optimally
with the proton $1h_{11/2}$  to favor a 
prolate shape. On the oblate side, the interaction 
between the relevant orbitals, leading to 
the corresponding  minima, takes place around 
$Q_{20} =-5$ b. The final net 
effect of  these driving forces is the
appearance of   prolate and oblate minima  
that become saddle points inside the 
{\it{island of triaxiality}} 
(see 
also the
discussion
below).

Below $^{182}$Pt, our axially symmetric 
calculations predict strong prolate deformations.
In this case there is a strong gap on the neutron  sector
centered at $Q_{20}$= 10 b, as it can be  seen
from the neutron SPE plot of $^{180}$Pt.
The most active deformed orbitals are the
ones coming from the neutron $1i_{13/2}$
and  $2f_{7/2}$. They interact 
with the deformed proton orbitals coming from the   
$1h_{11/2}$ and $2d_{5/2}$
to produce strongly prolate deformed shapes. 
The secondary oblate minima predicted for 
these nuclei, can be associated with 
the proton and neutron energy gaps observed 
around $Q_{20}$= -5 b. Further down in neutron
number, the neutron Fermi level ${\lambda}_{N}$ explores 
regions lower in energy resulting 
in  less pronounced prolate minima located 
around $Q_{20}$= 4 b in 
the neutron deficient isotopes
$^{166,168}$Pt. 

Let us now turn our attention to the origin of triaxiality
and for this, proton and neutron SPE plots
are shown in Figs. \ref{speAXTRI-proton} and 
\ref{speAXTRI-neutron}, respectively, as functions of the 
triaxial deformation parameter $\gamma$
\cite{Heenen-nature,ours2} for the nucleus $^{188}$Pt. 
We consider $Q=$7.8 b which corresponds to a rather large
region in the $Q-\gamma$ plane near the triaxial minimum (located
at $Q=$6.2 b, see left panel of Fig.\ref{figevolQ}) where the PES is very flat
in the two coordinates (see Fig. \ref{fig_d1s}). The value of $Q=$7.8 b also
corresponds to the position of the axial prolate minimum (see 
Fig. \ref{fig_axial}).
These plots allow us to identify
the $K$ values of the triaxial SPEs at the prolate 
($\gamma = 0^\circ$) and oblate ($\gamma = 60^\circ)$  limits
and the change of the $K$ contents observed in most of the levels
as $\gamma$ evolves. Typical examples in this context, are the
negative parity K=1/2 proton level with SPE $\epsilon \approx $ -3 MeV
at $Q$=7.8 b and $\gamma = 0^\circ$ which transforms into a 
$K=9/2$ level at $\gamma = 60^\circ$ and the 
positive parity $K=13/2$ neutron  level
with SPE $\epsilon \approx $ -4.2 MeV at $Q$=7.8 b and $\gamma = 0^\circ$
which transforms into a 
$K=1/2$ level at $\gamma = 60^\circ$. The rather 
low level density below the proton 
Fermi level for  $\gamma$ between $0^\circ$  and  $30^\circ$  favors
the flateness of the energy curve as a function of $\gamma$ helping thereby
the development of the triaxial minimum in $^{188}$Pt around 
$\gamma = 30^\circ$. From this plot we also conclude that
the isotopes with two protons less (Os isotopes) will be more prone to
triaxiallity as discussed in Ref \cite{ours2}. Concerning the 
behaviour of the neutron SPEs  with $\gamma$ depicted on Fig.
\ref{speAXTRI-neutron} we observe a region of low level density 
in the interval of $\gamma$ between $0^\circ$  and  $20^\circ$  that 
would favor the development of triaxiallity. From there on the level
density increases and therefore, in order to develop a triaxial
minimum, the system is forced to change its $Q$ deformation 
to lower values (see Fig. \ref{fig_d1s} for the nucleus  under 
consideration, $^{188}$Pt). The removal of two or four neutrons
makes the level density in the range of $\gamma$ between 
$0^\circ$  and  $20^\circ$
even lower than in the case of $^{188}$Pt explaining the $\gamma$ 
deformed minima observed in Fig. \ref{evol-min-gamma} for the nuclei 
$^{184,186}$Pt. On the other hand, the addition of two or four extra
neutrons leads to a decrease of the level density in the interval
between $\gamma \approx  30^\circ$ and 
$\gamma =  60^\circ$
 favoring
the appeareance of triaxial minima in $^{190,192}$Pt and flat curves
in $^{194,196}$Pt, as observed in Fig. \ref{evol-min-gamma}.

In the spirit of the shell model, it is also interesting 
to compute the  spherical occupancies
$\nu (lj,Q,\gamma)$ of the different  $lj$ orbitals in the 
(usually) deformed ground states 
$| \Phi_{\rm HFB}(Q,\gamma) \rangle$ 
of the Pt isotopes studied in this paper. They are given by
\begin{small}
\begin{eqnarray} \label{eq-occup}
\nu (lj,Q,\gamma) = 
\sum_{n} \sum_{m} \langle \Phi_{\rm HFB}(Q,\gamma) | c^+_{nljm} c_{nljm} | 
\Phi_{\rm HFB}(Q,\gamma) \rangle 
\end{eqnarray}
\end{small}
where $c^+_{nljm}$ and $c_{nljm}$ are 
the creation and annihilation operators of spherical harmonic oscillator orbits 
characterized by the
quantum numbers $n$, $l$, $j$, $m$. The sum in $m$ is introduced to make the 
quantity (\ref{eq-occup}) invariant under changes in 
orientation (and therefore to represent a genuine
"spherical" quantity). The sum in the radial quantum 
number $n$ does not allow
to pin down which specific HO orbital is occupied but, on the other 
hand, allows us to get rid
of the uncertainties associated to the fact that the radial wave function of 
the nuclear orbitals is close but not exactly the one of the HO.

The proton spherical occupancies in the ground state wave functions
of all the Pt isotopes considered are shown on  the left 
panel of Fig. \ref{occupations}. As the number of protons
remains constant along the isotopic chain one could expect a flat 
behavior of all the occupancies. However, we observe in that figure
how the occupancies of the different orbitals can be classified in two
different regimes, namely the weak deformation regime including $^{166-172}$Pt
and $^{188-204}$Pt and the strong deformation 
regime including the nuclei
$^{174-186}$Pt. It is noteworthy that in each of the regimes the proton
occupancies remain rather constant irrespective of the $\gamma$ deformation and
even the specific value of the $Q$ deformation parameter. The strong
deformation regime differs from the weak deformation one in that the 
$d_{5/2}$, $d_{3/2}$ and  $h_{11/2}$ orbitals loose occupancy in favor
of the $f_{7/2}$, $f_{5/2}$ and $h_{9/2}$. This rearrangement of the occupancies
is mainly due to the smearing out of the Fermi surface as a consequence of the
increasing proton pairing correlations (see Fig. \ref{figevolQ}) and 
to a lesser extent to the quadrupole
interaction among orbits that can transfer particles from one orbit to another.
Given the shift of occupancies from the $d_{5/2}$, $d_{3/2}$ and  $h_{11/2}$ 
to the $f_{7/2}$, $f_{5/2}$ and $h_{9/2}$ orbitals
we can interpret the well
deformed ground state of $^{174-186}$Pt as a multiparticle-multihole
excitation out of a  reference spherical ground state. 
The number of
particles exchanged in this kind of spherical shell model language is
in between two to four according to the results shown in the left panel of
Fig. \ref{occupations}.

In the case of neutrons, shown on  the right 
panel of Fig. \ref{occupations}, as neutron number increases we are occupying orbitals
belonging to the $N=5$ negative parity major shell and the positive parity
intruder $i_{13/2}$. As a consequence of deformation, the spherical orbitals
are mixed up and therefore placing two particles in a given deformed orbital
by no means implies placing two particles in an spherical orbit. The two particles
will be distributed among all the components of the deformed orbital when expressed
in the spherical basis. As a consequence, the behavior of the spherical 
occupancies with neutron number is more or less linear in the whole interval
and for all the orbitals involved. The only noticeable deviation from this trend
takes place when entering the strong deformation regime where the $f_{7/2}$,
$h_{9/2}$ and $h_{11/2}$ orbitals loose particles in favor of the
high $j$ orbitals $i_{13/2}$, $i_{11/2}$, $j_{13/2}$ and $j_{15/2}$. Also the
$g_{9/2}$ orbital gets more particles (through the coupling to the $N=6$ orbital).
This change in occupancies can be mostly attributed to the quenching of neutron
pairing correlations in the ground state of the strongly deformed isotopes  $^{174-186}$Pt.
When the weak deformation regime is entered at $A=188$ a much smoother behavior
of the spherical occupancies is recovered. This, together with the smooth behavior
of proton occupancies is quite unexpected as in the weak deformation regime
the different isotopes have a variety of ground state deformations ranging from
prolate to triaxial to oblate. The conclusion is that the occupancies 
are more sensitive to the magnitude of the deformation $Q$ than to the
$\gamma$ degree of freedom. 

The proton levels closest to the Fermi level are the $s_{1/2}$, $h_{11/2}$, $d_{3/2}$ 
and $d_{5/2}$ and therefore are the ones expected to strongly interact with the
neutron spin orbit partner according to the FP mechanism. The more effective orbitals
are those with high $j$ and therefore the proton $h_{11/2}$ is expected to strongly
interact with neutrons $h_{9/2}$ which is empty at the beginning of the chain and gets
steadily occupied as more neutrons are added. Also a strong interaction is expected with
the neutron $i_{13/2}$ orbital that shares with the $h_{9/2}$ one the occupation 
pattern as a function of $N$. It is also noteworthy to point out how proton $h_{9/2}$, that
should be empty according to the spherical shell model, gets some occupancy for the
well deformed nuclei $^{174-186}$Pt as a consequence of the smearing out of the 
Fermi surface, that allows those orbitals to interact via FP with
the neutrons $h_{11/2}$ and $g_{9/2}$. It could be argued that the proton $h_{11/2}$ 
orbital is loosing two particles in the strong deformation regime and this could imply
a strong impact in its interaction with its neutron spin-orbit partner. This would
be true if the $h_{11/2}$ orbital were occupied by a small amount of neutrons (of the
order of two) as this would imply completely emptying out the orbtial,  but it has to
be bear in mind that there are ten neutrons at the begining of the region under consideration.

\section{Conclusions}
\label{conclusions}

In this paper we have studied the evolution of the ground state 
nuclear shapes  in a series 
of Pt isotopes ranging from $N=88\, (A=166)$ up 
to $N=126\, (A=204)$, covering practically one complete major shell.
The study has been performed within 
the selfconsistent HFB approach based on the 
D1S \cite{gogny-d1s} and the recent D1N \cite{gogny-d1n} and D1M 
\cite{gogny-d1m}   parametrizations of the Gogny interaction
and we have  included, in addition to the axially symmetric limit, the 
triaxial degrees of freedom $\beta$ and 
$\gamma$.

From the analysis of the axially symmetric limit, we conclude that a 
sudden prolate to oblate shape change occurs around $N=110\, (A=188)$
for all the Gogny parametrizations considered. This result is also in 
agreement with those obtained either with Skyrme forces or from
relativistic mean field calculations
\cite{Robledo-2,Sharma-Ring,ours2,Sauvage}. On the other hand, when 
triaxiality is taken into account, the picture that 
finally emerges is that of smooth transitions between 
the different shape regimes. We find that the absolute minimum of the 
PESs for the Pt isotopes evolves from prolate shapes
with increasing values of their quadrupole moments $Q$ in the lighter 
isotopes $A=166-182$ (with the exception of  $A=172-176$, which 
exhibit a tendency to triaxiality), to triaxial $\gamma$-soft in the 
intermediate isotopes with $A=184-196$, and to oblate shapes in the 
most neutron-rich isotopes $A=198-202$. Finally, the  isotope
$^{204}$Pt becomes spherical. By analyzing
PECs and the $Q-\gamma$ landscapes, we observe 
that the (axial) prolate and oblate minima, well separated 
by high energy barriers in the $\beta$ degree of freedom, are 
softly linked along  the $\gamma$ direction. Indeed, most 
of the secondary axial minima become saddle 
points when the $\gamma$ degree of freedom is included
in the analysis. Pairing energies and Thouless-Valatin 
moments of inertia have also been analyzed 
as functions of the number of neutrons, finding some correlation 
with the evolution of the quadrupole deformation $Q$. Such a 
correlation, on the 
other hand, is not so evident in the case of the $\gamma$ angle.

We consider the similarity between the mean field  
ground state properties predicted for the considered 
Pt chain with the most recent versions of the Gogny interaction 
(i.e., D1S, D1N  and D1M)
employed
in the present study  as very positive.
On  one hand, the results give us confidence
concerning the  robustness  of the predictions 
against the details of the particular effective Gogny interaction
used. The robustness is reinforced when
the trend of our calculations is compared with the ones obtained 
with Skyrme forces \cite{Robledo-2,ours2} and other theoretical approaches
\cite{review,bengtsson,Cederwall-exp,Ce-other}, which are
 again very similar. On the other hand, our results also 
point to the fact that the new incarnations D1N and D1M 
 of the  Gogny interaction, based on fitting protocols
more in the direction of astrophysical 
applications \cite{gogny-d1n,gogny-d1m}, essentially
keep the predictive power of the Gogny-D1S force \cite{gogny-d1s} already
considered as a (standard) global force.

We have analyzed the proton and neutron SPEs as  functions
of both axial and triaxial deformation parameters in some
illustrative examples. The analysis has been 
done in terms of the density of levels around the Fermi surface
(Jahn-Teller effect \cite{JTE-1})
and the Federman-Pittel mechanism \cite{pittel}. Our discussion has
also been illustrated with the calculation of
the spherical occupancies
in the ground state wave functions of the considered Pt nuclei. As 
a result, we obtain a qualitative 
understanding of the emergence of deformed configurations
 with 
two main ingredients which are, the 
energy gaps that appear at different deformations in the SPEs of 
neutrons
when the Fermi level ${\lambda}_{N}$ crosses different regions, and
the special role of the overlap between the proton $ 1h_{11/2}$ and the 
neutron $ 1i_{13/2}$ orbitals.

The study of the low-lying excitation spectra and transition rates
in conjunction with shape transitions, would require to extend the
present mean field approach to take into account
correlations related to restoration of broken symmetries
and fluctuations of collective variables. The restoration of
broken symmetries would imply, among others, triaxial projections to
restore rotational symmetry and this is a very difficult and delicated
issue with realistic forces
\cite{Bender.08,Duguet-kernels-1,Duguet-kernels-2,Duguet-kernels-3}. 
These difficulties 
lead to consider instead of the exact projection some kind of approximation
to it that could eventually end up in a kind of  collective Bohr 
Hamiltonian 
\cite{bertsch,Libert-1,GOA-other-1,Prochniak}
with deformation dependent parameters. 
Upon completion of this work, a preprint has appeared 
\cite{delaroche10} dealing with the calculation
of $2^+$ excitation energies in the framework of the 5-D Bohr hamiltonian
method with parameters extracted from a microscopic mean field calculation
with the Gogny D1S force. As a consequence of the number of nuclei considered
(around two thousand) in that calculation, their analysis of the mean field
results is not as exahustive as ours.

\begin{acknowledgments}
This work was partly supported by  MEC
(Spain) under Contract FPA2007-66069, MICINN (Spain) under Contract
FIS2008-01301, the Consolider-Ingenio 
2010 program CPAN (Spain) under Contract CSD2007-00042 and Junta de
Andaluc\'ia (Spain)
under Contract P07-FQM-02962. The first steps of this work were
undertaken in the framework of the FIDIPRO program (Academy of 
Finland and University of Jyv\"askyl\"a) and one of us (R.R)
thanks Profs. J. Dobaczewski, J. \"Aysto, R. Julin and the experimental 
teams of the University of Jyv\"askyl\"a (Finland) for 
warm hospitality and encouraging discussions. We thank
Prof. S. Goriely (Universit\'e Libre de Bruxelles, Belgium)                                                 
for making the parametrization Gogny-D1M available to us 
prior to publication and also for valuable 
suggestions. Valuable 
suggestions from Prof. K. Heyde are also acknowledged.

\end{acknowledgments}

\end{document}